\pdfoutput=1 
\documentclass[aps,prd,twocolumn,superscriptaddress,amsfont,graphicx,nofootinbib,preprintnumbers]{revtex4}

\usepackage{color,graphicx,epsfig}
\usepackage{ifpdf}
\usepackage{amsmath}
\usepackage{bm}
\usepackage{color}
\usepackage[english]{babel}
\usepackage{graphicx}
\usepackage{amsfonts}
\usepackage{amssymb}
\usepackage{braket}
\usepackage{hyperref}
\usepackage{enumerate}
\usepackage{subfigure}

\definecolor{nicered}{rgb}{0.7,0.1,0.1}
\definecolor{nicegreen}{rgb}{0.1,0.5,0.1}
\hypersetup{colorlinks,citecolor= nicegreen,linkcolor= nicered}

\newcommand{\as}{\alpha_s}
\newcommand{\aso}{\bar\alpha_{s}}
\newcommand{\asb}{\alpha_{s,b}}
\newcommand{\asbo}{\bar\alpha_{s,b}}
\newcommand{\ep}{\epsilon}
\newcommand{\slashed}{\slash \hspace{-0.19cm}}

\newcommand{\be}{\begin{equation}}
\newcommand{\ee}{\end{equation}}
\newcommand{\bea}{\begin{eqnarray}}
\newcommand{\eea}{\end{eqnarray}}

\definecolor{Red}{rgb}{1.,0.,0.}
\definecolor{randomcolour}{rgb}{0.2,0.5,0.7}

\DeclareMathAlphabet\mathbfcal{OMS}{cmsy}{b}{n}

\def\OMIT#1{}

\definecolor{darkred}{rgb}{0.9,0,0}

\definecolor{darkgreen}{rgb}{0,0,0.9}

\definecolor{darkblue}{rgb}{0,0,0.9}

\allowdisplaybreaks[1]

\begin{document}

\title{Three-loop helicity amplitudes for photon+jet production}

\author{Piotr Bargie\l{}a}
\email{piotr.bargiela@physics.ox.ac.uk}
\affiliation{Rudolf Peierls Centre for Theoretical Physics, University of Oxford, Clarendon Laboratory, Parks Road, Oxford OX1 3PU, U.K.}
\author{Amlan Chakraborty}
\email{amlancha@buffalo.edu}
\affiliation{University at Buffalo, The State University of New York, Buffalo 14260, USA}
\author{Giulio Gambuti}
\email{giulio.gambuti@physics.ox.ac.uk}
\affiliation{Rudolf Peierls Centre for Theoretical Physics, University of Oxford, Clarendon Laboratory, Parks Road, Oxford OX1 3PU, U.K.}

\preprint{
	OUTP-22-14P,MSUHEP-22-039
}

\begin{abstract}
We present three-loop helicity amplitudes for the production of a single photon in association with one jet in Quantum Chromodynamics, a final state which provides a standard candle of the Standard Model at the Large Hadron Collider.
We employ a recently-proposed variation of the so-called tensor projection method in the 't Hooft-Veltman scheme (tHV) which avoids the computation of contributions due to unphysical ($-2 \epsilon$)-dimensional polarisations of the external states.
We obtain compact analytic results expressed in terms of harmonic polylogarithms.
\end{abstract}

\maketitle 

\allowdisplaybreaks

\section{Introduction}
\label{sec:intro}

Scattering amplitudes are one of the central quantities in Quantum Field Theory. In addition to the intrinsic beauty of their mathematical structure, they provide a bridge between theory and experiment. High-precision amplitudes in Quantum Chromodynamics (QCD) are essential to compute accurate theoretical predictions that, together with increasingly precise measurements from colliders, allows for a scrutiny of the structure of the Standard Model, and for constraining New Physics models. 

It is well known that the number of external particles and of internal loops greatly influences the complexity of amplitude calculations. Up until a few years ago, the state of the art for massless four-particle scattering was the three-loop four-gluon amplitude in $\mathcal{N}=4$ Super-Yang-Mills~\cite{Henn:2016jdu}, where calculations are simplified by the high amount of symmetry in the theory. The same amplitude was also computed in the planar limit of pure Yang-Mills in Ref.~\cite{Jin:2019nya}. 
In QCD similar three-loop calculations proved until recently to be too computationally prohibitive to be performed due to the lack of symmetry.

The first analytic results for a QCD three-loop four-point amplitude were obtained for the colour singlet process $q \bar q \to \gamma \gamma$ in Ref.~\cite{Caola:2020xup}. Building on this, the more computationally involved colour singlet production $g g \to \gamma \gamma$ was computed in Ref.~\cite{Bargiela:2021wuy}. Finally, all amplitudes involving four coloured partons, $i.e.$ $g g \to g g $, $q \bar{q} \to q \bar{q}$, $q \bar{q} \to g g $ and all possible crossings of external states, were obtained in Refs.~\cite{Caola:2021izf,Caola:2021rqz,Caola:2022dfa}.
In these processes, starting a three loops, there start appearing new contributions to the structure of infrared (IR) divergences. These are associated with the exchange of colour charge between all four external legs through the emission and absorption of soft gluons. This is referred to as \textit{quadrupole radiation} and it increases the complexity of the corresponding amplitudes.

In this work we tackle the last three-loop four-point massless amplitudes in QCD involving partonic initial states:
$g g \to g \gamma$ and $q \bar{q} \to g \gamma$.

Phenomenologically, this amplitude is relevant for the $pp\to\gamma+j$ process, i.e. direct photon production with a reconstructed jet.
It is one of the standard candles of the Standard Model at the Large Hadron Collider (LHC).
Theoretical QCD predictions for this process exist at the next-to-next-to-leading order (NNLO)~\cite{Campbell:2016lzl}.
Providing even higher order corrections would lead to a more precise comparison with the LHC data, which is important for current and especially future LHC runs~\cite{Campbell:2016lzl}.
The three-loop amplitudes in quark-pair and quark+gluon initiated channels contribute to the N3LO cross section.
The three-loop gluon-pair initiated amplitude starts contributing only at N4LO, however it is enhanced by the Parton Distribution Function (PDF) of the gluon, which may at least partially compensate for the strong coupling suppression.

For photon+jet production at hadron colliders, we consider the two independent partonic channels
\begin{equation} \label{eq:processes}
\begin{split}
& g (p_1) + g(p_2)  \to g(-p_3) + \gamma(-p_4) \,, \\
& q (p_1) + \bar{q}(p_2)  \to g(-p_3) + \gamma(-p_4) \,.
\end{split}
\end{equation}
The remaining $q g \to q \gamma$ and $ \bar{q} g  \to \bar{q} \gamma$ channels can be obtained via crossing of $q \bar{q} \to g \gamma$.
We treat all four-momenta as incoming and massless
\begin{equation}
p_1+p_2+p_3+p_4=0\,, \qquad p_i^2=0 .
\end{equation}
The kinematic Mandelstam invariants of the process,
\begin{equation}
s = (p_1+ p_2)^2\,,~ t = (p_1+p_3)^2\,,~u = (p_2+p_3)^2\,,
\end{equation}
are related by momentum conservation $s+t+u = 0$.
Since the overall mass dimension of the amplitude is fixed, the non-trivial kinematic dependence can be expressed in terms of a single dimensionless ratio
\begin{equation}
x = -\frac{t}{s}\,.
\end{equation}
On the physical Riemann sheet, we have~\footnote{Technically, Eq.~\eqref{eq:anCont} is imprecise since the condition $s+t+u = 0$ has to be always satisfied. This makes analytic continuation for massless $2\to2$ scattering delicate, see e.g. Ref.~\cite{Caola:2021rqz} for a discussion in the context of three-loop amplitudes.}
\begin{equation}
\label{eq:anCont}
s>0\,, t<0\,, u<0\,, \quad s_{ij} \to s_{ij}+i\delta\,,
\end{equation}
where $s_{ij} = 2p_i\cdot p_j$.
\\

This paper is organised as follows: in Section~\ref{sec:comp} we describe the colour and Lorentz space decomposition of the amplitudes. The definition of the helicity amplitudes is given in Section~\ref{sec:hel} where we also fix our notation within the spinor helicity formalism and describe the workflow used for this computation. Section~\ref{sec:uvir} describes renormalisation and IR subtraction of the helicity amplitudes. Finally, we give more details on the results in Section~\ref{sec:results} and provide concluding remarks in Section~\ref{sec:concl}.
\section{Colour and tensor structures}
\label{sec:comp}
For both processes in Eq.~\eqref{eq:processes}
we can collect an overall colour factor $\mathcal{C}$ in front of the amplitude:
\begin{align}
\mathcal{A} = \mathcal{C} \, A \, ,
\end{align}
where 
\begin{align}
\mathcal{C}  =
    \begin{cases}
         \text{Tr}(T^{a_1}\!T^{a_2}\!T^{a_3}\!) - (2 \longleftrightarrow 3), & \text{for } g g \to g \gamma, \\[8pt]
        T^{a_3}_{i_1 i_2}, & \text{for } q \bar{q} \to g \gamma.
    \end{cases}
\end{align}
Above $i_n(a_n)$ refers to a $SU(N_c)$ index in the fundamental(adjoint) representation
and $T^{a}$ are the fundamental generators of $SU(N_c)$, normalised such that $\text{Tr}(T^{a}T^{b}) = \frac{1}{2} \delta^{a_1 a_2}$.
The colour stripped amplitude $A$ depends on the number of active quark flavours $n_f$, the electric coupling of the different quark flavours $Q_f$ and the  fermionic loop factor
\begin{align}
n_f^{(V)} = \sum_{f=1}^{n_f} Q_f\,.
\end{align}%
After extracting all colour structures, $A$ can be further decomposed onto a basis of $n_t$ independent Lorentz tensor structures
\begin{equation}
A = \sum_{i=1}^{n_t} T_i \mathcal{F}_i \,.
\label{eq:ampoT}  
\end{equation}
We work in the tHV regularisation scheme, where internal states are in $d$ dimensions but the external momenta and polarisations are kept in 4 dimensions. In this scheme, we follow a recent proposal~\cite{Peraro:2019cjj,Peraro:2020sfm} that allows one us to remove the irrelevant $(-2 \epsilon)$-dimensional external helicity states and to work with a set of tensors $T_i$ whose number coincides with that of the independent helicity configurations.\\
In the $g g \to g \gamma$ channel, $n_t=8$ and with the cyclic gauge $\epsilon_i \cdot p_{i+1} = 0$ and $p_5 \equiv p_1$ we find 
\begin{align}
\label{eq:tengg}
T_1 &= p_1\!\cdot\!\ep_2 \; p_1\!\cdot\!\ep_3 \; p_2\!\cdot\!\ep_4 \; p_3\!\cdot\!\ep_1, \notag\\
T_2 &= \ep_3\!\cdot\!\ep_4 \; p_1\!\cdot\!\ep_2 \; p_3\!\cdot\!\ep_1 , \quad
T_3 = \ep_2\!\cdot\!\ep_4 \; p_1\!\cdot\!\ep_3 \; p_3\!\cdot\!\ep_1 , \notag\\
T_4 &= \ep_2\!\cdot\!\ep_3 \; p_2\!\cdot\!\ep_4 \; p_3\!\cdot\!\ep_1 , \quad
T_5 = \ep_1\!\cdot\!\ep_4 \; p_1\!\cdot\!\ep_2 \; p_1\!\cdot\!\ep_3 , \notag\\
T_6 &= \ep_1\!\cdot\!\ep_3 \; p_1\!\cdot\!\ep_2 \; p_2\!\cdot\!\ep_4 , \quad
T_7 = \ep_1\!\cdot\!\ep_2 \; p_1\!\cdot\!\ep_3 \; p_2\!\cdot\!\ep_4 , \notag\\
T_8 &= \ep_1\!\cdot\!\ep_2 \; \ep_3\!\cdot\!\ep_4 + \ep_1\!\cdot\!\ep_4 \; \ep_2\!\cdot\!\ep_3 + \ep_1\!\cdot\!\ep_3 \; \ep_2\!\cdot\!\ep_4 .
\end{align}
In the $q \bar{q} \to g \gamma$ channel, $n_t=4$ and with the gauge choice $\ep_3 \cdot p_2 = \ep_4 \cdot p_1 = 0$ we get
\begin{align}
\label{eq:tenqq}
T_1 &= \bar{u}(p_2) \slashed{\ep_3} u(p_1) \: \ep_4 \!\cdot\! p_2, \notag\\
T_2 &= \bar{u}(p_2) \slashed{\ep_3} u(p_1) \: \ep_4 \!\cdot\! p_1, \notag\\
T_3 &= \bar{u}(p_2) \slashed{p_3} u(p_1) \: \ep_3 \!\cdot\! p_1 \, \ep_4 \!\cdot\! p_2, \notag\\
T_4 &= \bar{u}(p_2) \slashed{p_3} u(p_1) \: \ep_3 \!\cdot\! \ep_4\, .
\end{align}
The form factors $\mathcal{F}_i$ can be extracted from $A$ with appropriate projectors, defined such that $\sum_{pol} P_j T_i = \delta_{ji}$, see e.g. Refs.~\cite{Peraro:2019cjj,Peraro:2020sfm} for the full discussion.

\section{Helicity amplitudes}
\label{sec:hel}

In order to obtain the helicity amplitudes $A_{\vec{\lambda}}$, it is enough to evaluate the tensors $T_i$ for fixed-helicity configurations ${\vec{\lambda}}$.
This is equivalent to a simple change of basis, and the helicity amplitude for the helicity configuration ${\vec{\lambda}} = \{\lambda_1,\lambda_2,\lambda_3,\lambda_4\}$ can be written as a linear combination of form factors $\mathcal{F}_i$
\begin{equation}
A_{\vec{\lambda}}
= \sum_{i=1}^{n_t} T_{i,{\vec{\lambda}}} \mathcal{F}_i
= \mathcal{S}_{\vec{\lambda}} \, \mathcal{H}_{\vec{\lambda}}\,.
\end{equation}
The overall spinor factors $\mathcal{S}_{\vec{\lambda}}$ can be extracted from $A_{\vec{\lambda}}$ using the spinor-helicity formalism, see e.g. Ref.~\cite{Dixon:1996wi} for a pedagogical introduction.
In this notation, fixed-helicity external quarks are defined as
\begin{equation}
| p \rangle = \overline{[ p |} = \frac{1+\gamma_5}{2} u(p)\,, \quad
| p ] = \overline{\langle p |} = \frac{1-\gamma_5}{2} u(p)\,,
\end{equation}
with $[ p | = \overline{u}(p) \frac{1-\gamma_5}{2}$ and
$\langle p | = u(p) \frac{1+\gamma_5}{2}$
treating particles and anti-particles on the same footing,
while polarisation vectors take the following form
\begin{equation}
\epsilon^\mu_{j,+}(p_j) = \frac{\langle p_j | \gamma^\mu | q_j ] }{ \sqrt{2} [ p_j q_j ]}\,, \quad
\epsilon^\mu_{j,-}(p_j) = \frac{\langle q_j | \gamma^\mu | p_j ] }{ \sqrt{2} \langle q_j p_j \rangle }\,,
\end{equation}
where $q_i$ is the massless reference vector corresponding to the $i$-th external gluon and is chosen consistently with the gauge conditions used to determine the tensor bases of Eqs.~\eqref{eq:tengg} and~\eqref{eq:tenqq}.  
For the $g g \to g \gamma$ channel we have $q_i = p_{i+1}$, where we identify $p_5 \equiv p_1$ and we choose the spinor factors to be
\begin{align}
\mathcal{S}_{++++} &= \frac{\langle1 2\rangle\langle3 4\rangle}{[ 1 2] [ 34] } \,, & 
\mathcal{S}_{-+++} &=
\frac{[ 1 2] [ 1 4] \langle2 4\rangle}{[ 3
	4] [ 2 3] [ 2 4] } \,,
\notag\\
\mathcal{S}_{+-++} &= \frac{[ 2 1] [ 2 4] \langle1
	4\rangle}{[ 3 4] [ 1 3] [ 1 4] } \,,&
\mathcal{S}_{++-+} &= \frac{[ 3 2] [ 3 4] \langle2
	4\rangle}{[ 1 4] [ 2 1] [ 2 4] } \,,
\notag\\
\mathcal{S}_{+++-} &= \frac{[ 4 2] [ 4 3] \langle2
	3\rangle}{[ 1 3] [ 2 1] [ 2 3] } \,,&
\mathcal{S}_{--++} &= \frac{[ 1 2] \langle3 4\rangle}{\langle1
	2\rangle[ 3 4] } \,,
\notag\\ 
\mathcal{S}_{-+-+} &= \frac{[ 1 3] \langle2 4\rangle}{\langle1 3\rangle[ 2
	4] } \,,&
\mathcal{S}_{+--+} &= \frac{[ 2 3] \langle1 4\rangle}{\langle2 3\rangle[ 1 4] } \,,&&
\label{eq:helamp}
\end{align}
while for the $q \bar{q} \to g \gamma$ channel we have $q_3 = p_2, \; q_4=p_1$ and define the spinor factors as
\begin{align}
\mathcal{S}_{-+--} &= \frac{2[3 4]^2}{\langle1 3\rangle[2 3]} \,, & 
\mathcal{S}_{-+-+} &= \frac{2\langle2 4\rangle[1 3]}{\langle2 3\rangle[2 4]} \,,
\notag\\
\mathcal{S}_{-++-} &= \frac{2\langle2 3\rangle[4 1]}{\langle2 4\rangle[3 2]} \,,&
\mathcal{S}_{-+++} &= \frac{2\langle3 4\rangle^2}{\langle3 1\rangle[2 3]} \,.
\label{eq:spin_weight_qqga}
\end{align}
The spinor-free helicity amplitude $\mathcal{H}_{\vec{\lambda}}$ can be expanded as a QCD perturbative series
\begin{align} \label{eq:helicity_pert}
\mathcal{H}_{\vec{\lambda}} &= 
\sqrt{4 \pi \alpha}\sqrt{4 \pi \asb}
\sum_{\ell=0}^{3} \left(\frac{\asb}{4\pi}\right)^{\ell} \mathcal{H}^{(\ell)}_{\vec{\lambda}}\,,
\end{align}
where we have factored out an overall electric coupling $e = \sqrt{4 \pi \alpha}$ as well as a bare strong coupling $g_{s,b} = \sqrt{4 \pi \asb}$.
$\mathcal{H}^{(\ell)}_{\vec{\lambda}}$ corresponds to the bare $\ell$-loop amplitude.
Note that since the $g g \to g \gamma$ channel is loop-induced, the first term in its perturbative expansion vanishes.
Conversely, the $q \bar{q} \to g \gamma$ channel contributes non-trivially at all four orders.
The main objective of this paper is to compute the three-loop term $\mathcal{H}^{(3)}_{\vec{\lambda}}$.
As a byproduct of our work, we have also recomputed all lower-loop amplitudes. 
We checked all one-loop helicity amplitudes numerically against \texttt{OpenLoops}~\cite{Cascioli:2011va,Buccioni:2019sur}. At the two-loop level, to the best of our knowledge, the only available analytic results are for the process $q \bar{q} \to g \gamma$ and are given in Ref.~\cite{Anastasiou:2002zn} in the form of a colour- and polarisation-summed interference with the tree-level, with which we found perfect agreement\footnote{The results presented in Ref.~\cite{Anastasiou:2002zn} are computed in the Conventional Dimensional Regularisation (CDR) scheme and with a IR subtraction scheme which differs from the one adopted in this paper. While the difference in dimensional regularisation scheme is immaterial at the level of the $O(\epsilon^0)$ part of the finite remainder provided by the authors of Ref.~\cite{Anastasiou:2002zn}, the difference in IR subtraction scheme is not. In order to bridge the gap, we performed a second subtraction of the IR poles of our two loop renormalised amplitudes using the definition of the finite part in Ref.~\cite{Catani:1998bh}, and combined all helicity amplitudes to obtain the same quantity computed in Ref.~\cite{Anastasiou:2002zn}. }. \\
We generate Feynman diagrams corresponding to each channel with \texttt{Qgraf}~\cite{Nogueira:1991ex}.
At three loops, there are 7356 graphs in the $g g \to g \gamma$ channel, and 5534 graphs in the $q \bar{q} \to g \gamma$ channel.
Then, we perform colour and Dirac algebra using \texttt{Form}~\cite{Vermaseren:2000nd}.
There are $\mathcal{O}(10^6)$ scalar Feynman integrals contributing to the form factors of each of the scattering process described in Eq.~\eqref{eq:processes}.
Since the integrals at hand are not all linearly independent, we can find relations between them.
Preliminarily, we exploited loop-momentum shift-invariance to reduce the complexity by a factor of about 20.
This is then followed by the most complicated step, which involves Integration-By-Parts (IBP) identities~\cite{Chetyrkin:1981qh} to relate the remaining integrals to a minimal independent basis set of Master Integrals (MIs).
In order to perform the IBP reduction, we have used the Laporta algorithm~\cite{Laporta:2000dsw} implemented in \texttt{Reduze 2}~\cite{Studerus:2009ye,vonManteuffel:2012np}, as well as in \texttt{Finred}~\cite{vonManteuffel:2014ixa}, which exploits syzygy-based techniques~\cite{Gluza:2010ws,Ita:2015tya,Larsen:2015ped,Bohm:2017qme,Schabinger:2011dz,Agarwal:2020dye}  and finite-field arithmetic~\cite{vonManteuffel:2014ixa,vonManteuffel:2016xki,Peraro:2016wsq,Peraro:2019svx}.
In this manner, we are left with 486 independent MIs.
They have been computed analytically as a series in the dimensional regulator $\ep=(4-d)/2$ in terms of Harmonic Polylogarithms (HPLs)~\cite{Remiddi:1999ew} in Ref.~\cite{Henn:2020lye} and later in Ref.~\cite{Bargiela:2021wuy}.
The evaluation of MIs is based on the differential equation approach applied to a canonical basis of master integrals~\cite{Henn:2013pwa}.
Substituting expressions for MIs leads to the final formula for the bare three-loop amplitude, expanded to $\mathcal{O}(\ep^0)$ with HPLs up to transcendental weight 6.

\begin{figure}
\centering
\subfigure[]{\includegraphics[width=.47 \textwidth]{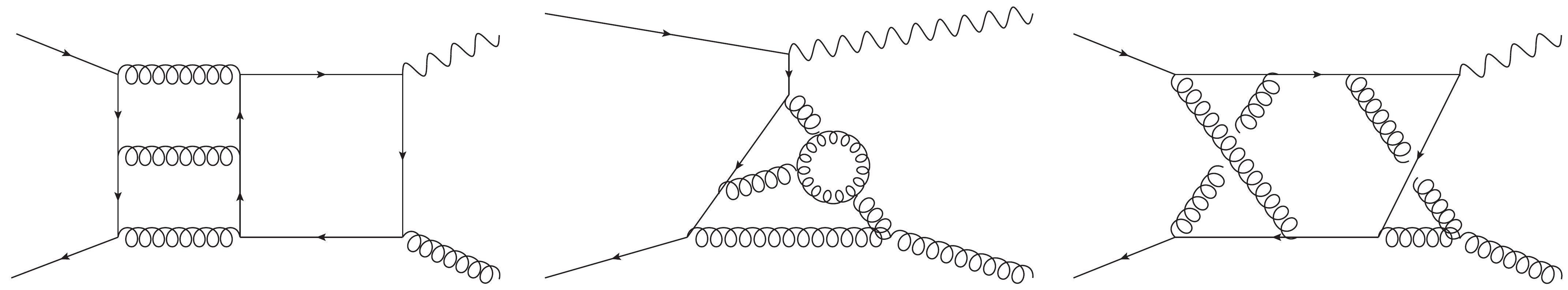}}
\end{figure}
\begin{figure}
\centering
\subfigure[]{\includegraphics[width=.47 \textwidth]{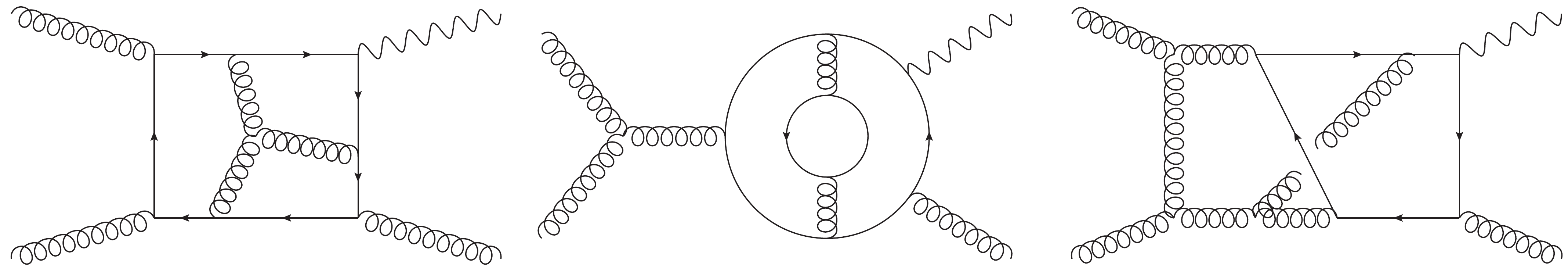}}
\caption{Sample three-loop diagrams for (a) the process $q \bar{q} \to g \gamma$ and (b) the process $gg \to g \gamma$. } \label{diagrams}
\end{figure}

\section{UV renormalisation and IR regularisation}
\label{sec:uvir}

The divergences appearing in the amplitudes treated in this paper are both of ultraviolet (UV) and infrared (IR) origin. When working in dimensional regularisation they are represented by poles in the dimensional regulator $\ep$.
By defining the $\overline{\text{MS}}$ renormalised strong coupling $\as(\mu)$ through the equation
\begin{equation}\label{eq:ren_coupling}
\asbo \mu_0^{2\ep} S_\ep = \aso(\mu)  \mu^{2\ep} Z\left[\aso(\mu) \right],
\end{equation}
one can obtain UV-finite amplitudes. Above, we have defined for convenience $\asbo = \asb/(4\pi)$ and
$\aso(\mu) = \as(\mu)/(4\pi)$.
The quantity $\mu$ is the renormalisation scale introduced in dimensional regularisation and $Z[\aso]$ reads
\begin{align}\label{eq:Zuv}
Z[\aso]  &=  1
- \aso  \frac{ \beta_0 }{\epsilon } +\aso^2 \left( \frac{\beta_0^2}{\epsilon^2} - \frac{\beta_1 }{2 \epsilon} \right)  \nonumber \\
&\quad
-\aso^3  \left( \frac{\beta_0^3}{\epsilon^3} - \frac{ 7}{6} \frac{\beta_0 \beta_1}{\epsilon^2}+ \frac{\beta_2}{3 \epsilon} \right)  + \mathcal{O}(\aso^4)\,.
\end{align}
The explicit form of the $\beta$-function coefficients $\beta_i$ can be found in ancillary files.
The perturbative contributions to the UV-renormalised helicity amplitudes $\mathcal{H}_{\vec{\lambda},\: \text{ren}}$ are obtained by expanding Eq.~\eqref{eq:helicity_pert} in $\aso(\mu)$. 

The poles in $\ep$ appearing in the renormalised amplitudes are of IR nature and their structure was described at two loops in~\cite{Catani:1998bh} and generalised to different processes~\cite{Sterman:2002qn,Aybat:2006wq,Aybat:2006mz} and to three loops in Refs.~\cite{Becher:2009cu,Becher:2009qa,Dixon:2009gx,Gardi:2009qi,Gardi:2009zv,Almelid:2015jia}. They assume a universal form across all massless gauge theories. Up to the three loop-order, one can write~\cite{Becher:2009cu,Becher:2009qa}
\begin{equation}\label{eq:IR_factorisation}
\mathcal{H}_{{\vec{\lambda}},\: \text{ren}} = \mathcal{Z}_{IR} \; \mathcal{H}_{{\vec{\lambda}},\: \text{fin}} \; ,
\end{equation}
where $ \mathcal{H}_{{\vec{\lambda}},\: \text{fin}} $ are \emph{finite
remainders} and $\mathcal Z_{IR}$ is in general a colour operator that acts on the colour structure of the amplitudes. 
It can be written in terms of the so-called soft anomalous dimension
$\mathbf\Gamma$ as
\begin{equation}\label{eq:exponentiation}
\mathcal{Z}_{IR} = \mathbb{P}\exp \left[ \int_\mu^\infty \frac{\mathrm{d} \mu'}{\mu'}  \mathbf{\Gamma}(\{p\},\mu')\right]  \; ,
\end{equation}
where the \textit{path ordering} operator $\mathbb{P}$ reorganises
colour operators in increasing values of $\mu'$ from left to
right and is immaterial up to three loops
since to this order $[\mathbf\Gamma(\mu),\mathbf\Gamma(\mu')] = 0$.
The soft
anomalous dimension can be written as
\begin{equation}\label{eq:dipole_+_quadrupole}
\mathbf{\Gamma}=  \mathbf{\Gamma}_{\text{dip}}  + \mathbf{\Delta}_4  \; .
\end{equation}
The \emph{dipole} term $\mathbf{\Gamma}_{\text{dip}}$ is due to the pairwise exchange of colour charge between external legs and reads
\begin{align}\label{eq:dipole}
\mathbf{\Gamma}_{\text{dip}}  &=  \sum_{1\leq i < j \leq 4} \mathbf{T}^a_i \; \mathbf{T}^a_j\; \gamma^\text{K} \; \ln{\left(\frac{\mu^2}{-s_{ij}-i \delta}\right)}  + \sum_{i=1}^4 \gamma^i \; ,
\end{align}
where $s_{ij} = 2p_i\cdot p_j$, $\gamma^{\text{K}}$ is the \textit{cusp anomalous dimension}
\cite{Korchemsky:1987wg,Moch:2004pa,Vogt:2004mw,Grozin:2014hna,Henn:2019swt,Huber:2019fxe,vonManteuffel:2020vjv} and $\gamma^{i=q,g}$ are the \textit{quark and gluon anomalous dimensions} \cite{Ravindran:2004mb,Moch:2005id,Moch:2005tm,Agarwal:2021zft}.
In Eq.~\eqref{eq:dipole} we have also introduced the standard colour insertion operators $\mathbf{T}^a_i$, which only act
on the $i$-th external colour index. Their action on the colour factors of our amplitudes is defined as follows:
\begin{alignat}{2}
(\mathbf{T}^a_i)_{b_i c_i} &=    - i {f ^a}_{b_i c_i} &~&\text{for a gluon}, \notag\\
(\mathbf{T}^a_i)_{i_i j_i} &=   + T^a_{i_i j_i} &&\text{for a (initial)final state (anti-)quark}, \notag \\
(\mathbf{T}^a_i)_{i_i j_i} &=   - T^a_{j_i i_i} &&\text{for a (final)initial state (anti-)quark}, \notag \\
\mathbf{T}^a_i &= 0&& \text{for a photon}. \label{eq:colour_insertions}
\end{alignat}
Performing the colour algebra with the definitions in Eq.~\eqref{eq:colour_insertions} we find the explicit value of the dipole anomalous dimensions in the two channels under consideration. Note that, since the amplitudes considered here feature a single colour structure, $\mathbf{\Gamma}$ acts by simple scalar multiplication. For $q\bar{q} \to g \gamma$ it reads
\begin{align}
\mathbf{\Gamma}^{q\bar{q} \to g \gamma}_{\text{dip}} &= \frac{1}{2} \! \bigg\{\! \frac{1}{N_c}\ln \!\left(\! \frac{\mu^2}{-s\!-\!i \delta}\! \right) + \\
 - N_c
 &\left[ 
\ln \!\left(\!  \frac{\mu^2}{-t\!-\!i \delta}\!\right) +
\ln \!\left(\!  \frac{\mu^2}{-u\!-\!i \delta}\!\right) 
\right] \! \bigg\} \gamma^\text{K} + 2 \gamma^q + \gamma^g, \notag
\end{align}
while for $g g \to g \gamma$ we get
\begin{align}
\mathbf{\Gamma}^{gg \to g \gamma}_{\text{dip}} &= -\frac{N_c}{2} \! \bigg\{\! \ln \!\left(\! \frac{\mu^2}{-s\!-\!i \delta}\! \right) + \\
 +
 & 
\ln \!\left( \! \frac{\mu^2}{-t\!-\!i \delta}\!\right) +
\ln \!\left(\!  \frac{\mu^2}{-u\!-\!i \delta}\!\right) 
\!\bigg\} \gamma^\text{K} + 3 \gamma^g.  \notag
\end{align}

The \emph{quadrupole} contribution $\mathbf{\Delta}_4$ in
Eq.~\eqref{eq:dipole_+_quadrupole} accounts instead for the exchange
of colour charge among (up to) four external legs and it appears for the first time at three loops, $ \mathbf{\Delta}_4
= \sum_{n=3}^\infty \aso^n \mathbf{\Delta}^{(n)}_4 $. 
Because the amplitudes considered in this paper feature only three coloured external states, $\mathbf{\Delta}_4$ assumes a simpler form compared to the full result given in Ref.~\cite{Almelid:2015jia}. In addition to this, the perturbative series for $gg \to g \gamma$ starts at one loop and therefore it receives no quadrupole correction at three loops.

For $q\bar{q} \to g \gamma$, the relevant contribution reads
\begin{align} \label{eq:quadrupole}
&\mathbf{\Delta}^{(3),q\bar{q} \to g \gamma}_4 =
- 24 N_c ( \zeta_5 + 2 \zeta_2 \zeta_3)\;.
\end{align}

We verified that the IR singularities of our three-loop amplitudes match perfectly those generated by 
Eqs.~\eqref{eq:IR_factorisation}-\eqref{eq:quadrupole},
which provides a highly non-trivial check of our results.

\section{Results}
\label{sec:results}
The expressions we obtained for the finite remainder $ \mathcal{H}_{{\vec{\lambda}},\: \text{fin}} $ are relatively compact, but still too long to be included in this manuscript. 
They are provided in computer-readable format
in the ancillary files accompanying the
\texttt{arXiv} submission of this manuscript.

\begin{figure}
    \includegraphics[width=1\linewidth]{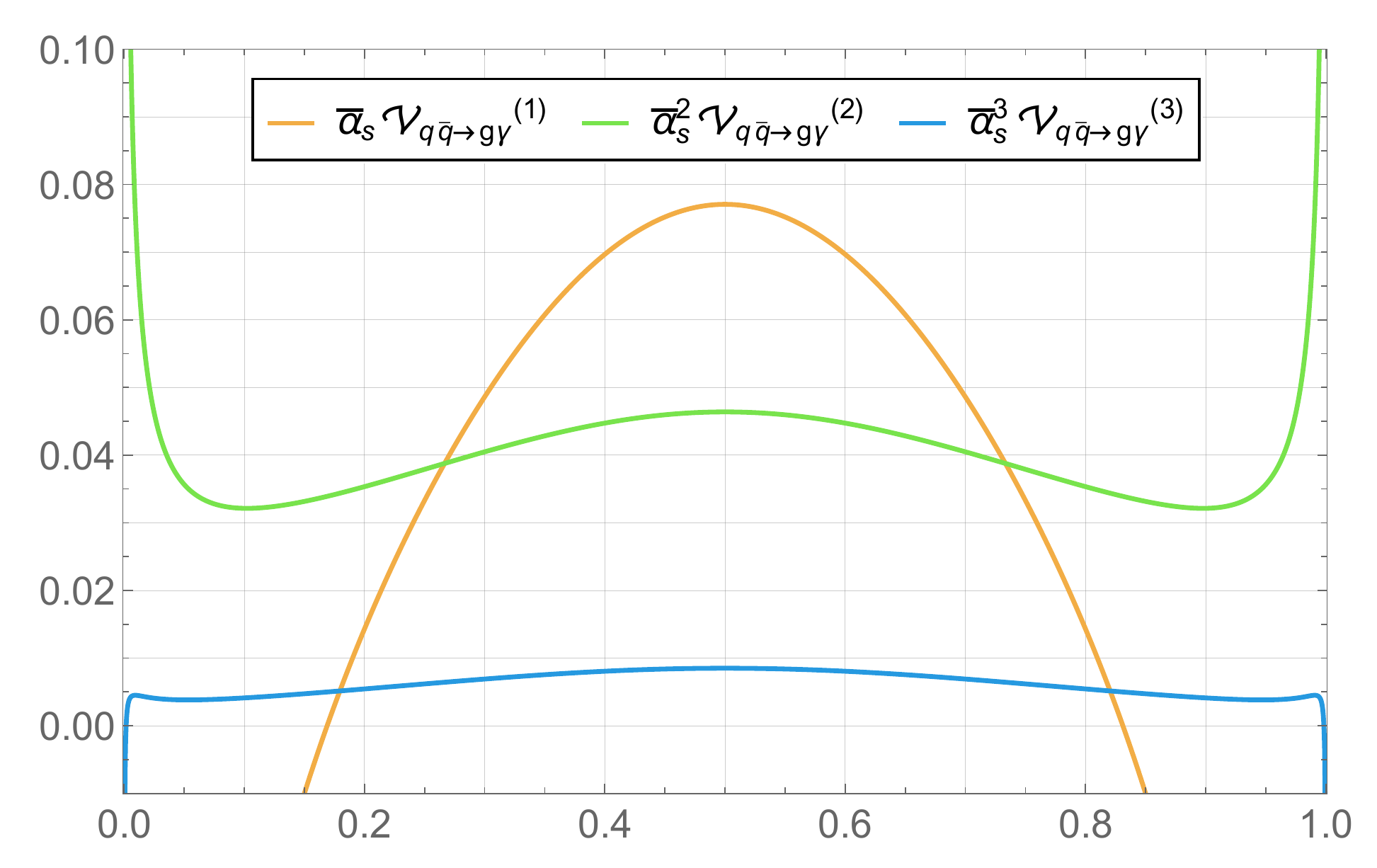}
    \caption{Perturbative expansion of the colour, helicity and flavour summed finite remainder of the amplitude squared for the process $q\bar{q}\to g \gamma$ as a function of $x=-t/s$. For simplicity we set $\as = 0.118$, $\mu^2 = s$, $N_c = 3$, $n_f = 5$ and $n_f^{(V)} = 1/3$.
}  \label{fig:full_amplitude}
\end{figure}

We also provide the explicit results for the channel $q g \to q \gamma$ which can be generated by crossing the $q \bar{q} \to g \gamma$ amplitude. The only other relevant channel for $pp \to j \gamma$ is $\bar{q} g \to \bar{q} \gamma$ and it can be obtained by charge conjugation of $q g \to q \gamma$, which leaves the helicity-stripped amplitudes unchanged.
When crossing the amplitudes, one has to permute Mandelstam invariants. In doing so, it is important not to cross more then one branch cut per transformation.
This can be guaranteed by an appropriate composition of multiple transformations.
All needed manipulations of HPLs can be performed with with \texttt{PolyLogTools}~\cite{Duhr:2019tlz} and the procedure is described in more detail in Ref.~\cite{Caola:2021rqz}.
It is also worth pointing out that when applying a crossing or a charge/parity transformation to the amplitudes associated to the processes in Eq.~\eqref{eq:processes}, one has to take care of applying the corresponding transformations to the spinor weights in Eqs.~\eqref{eq:helamp} and \eqref{eq:spin_weight_qqga}.

Finally, in order to showcase the fast and stable numerical evaluation of our amplitudes, in Figure~\ref{fig:full_amplitude} we provide a sample plot for the $q \bar{q} \to g \gamma$ channel, where we numerically evaluated the squared amplitude normalised to the tree-level. To define the quantities plotted in the figure more precisely, we first introduce the notation
\begin{align}
\langle \mathcal A^{(\ell)} | 
\mathcal A^{(\ell')}\rangle 
\equiv  
\sum_{f,\vec{\lambda},col}
\mathcal C^\dagger \mathcal C \;
|s_{\vec{\lambda}}|^2 \;
\mathcal H^{(\ell)^*}_{\vec{\lambda}, \rm fin}
\mathcal H^{(\ell')}_{\vec{\lambda}, \rm fin}
\end{align}
for the interference between two amplitudes summed over all internal quark flavours, all helicity configurations and all colours of the external states. 
With this, we can write
\begin{align}\label{eq:virtualdef}
    \mathcal{V}^{(1)} &=  \frac{  2 \:\text{Re}  \langle \mathcal A^{(0)} | \mathcal A^{(1)}\rangle }{\langle \mathcal A^{(0)} | \mathcal A^{(0)}\rangle } , \notag\\
    \mathcal{V}^{(2)} &=  \frac{   \langle \mathcal A^{(1)} | \mathcal A^{(1)}\rangle }{\langle \mathcal A^{(0)} | \mathcal A^{(0)}\rangle } + \frac{  2 \:\text{Re}  \langle \mathcal A^{(0)} | \mathcal A^{(2)}\rangle }{\langle \mathcal A^{(0)} | \mathcal A^{(0)}\rangle }, \notag\\  
    \mathcal{V}^{(3)} &=  \frac{   2 \:\text{Re}  \langle \mathcal A^{(1)} | \mathcal A^{(2)}\rangle }{\langle \mathcal A^{(0)} | \mathcal A^{(0)}\rangle } + \frac{  2 \:\text{Re}  \langle \mathcal A^{(0)} | \mathcal A^{(3)}\rangle }{\langle \mathcal A^{(0)} | \mathcal A^{(0)}\rangle },      
\end{align}
where the process dependence has been left implicit.
We point out that the quantities plotted in Figure~\ref{fig:full_amplitude} nicely show convergence of the perturbative series. However it should be kept in mind that they only represent the virtual contributions to the cross-section and depend on the IR subtraction scheme. 
In addition we observe the same alternating behaviour of the different perturbative orders when approaching the limits $x\to0$ and $x\to1$ that is also present in the $q\bar q \to g g $ and $gg \to g g $ channels \cite{Caola:2022dfa}.

\section{Conclusions}
\label{sec:concl}
In this paper we have completed the computation of the last three-loop four-point massless QCD scattering amplitudes with partons in the initial state. To do so,  we employed a refined version of the tensor projection method which considerably reduces the amount of calculations required.
At one and two loops, we have checked the consistency of our result against the available literature and found perfect agreement. 
At three loops, our amplitudes have the correct UV and IR structure, including the quadrupole contribution that appears at this perturbative order for the first time.
The corrections provided in this paper start contributing to the differential cross section at N3LO and in the future they will hopefully allow to achieve better precision on the prediction of photon+jet observables at hadron colliders.\\

{\bf Acknowledgements} We are grateful to A. von Manteuffel for providing the IBP relations through his private code \texttt{Finred}. We also thank F. Caola and L. Tancredi for insightful discussions and comments on this manuscript. AC is grateful to the Department  of  Physics  and  Astronomy,  Michigan  State  University for the hospitality for the time during which the initial part of the calculation was performed.
The research of PB was supported by the ERC Starting Grant 804394 HipQCD. AC was supported by the National Science Foundation by grants PHY-1652066 and NSF-PHY-2014021.
GG was supported by the Royal Society grant URF/R1/191125.

\bibliographystyle{apsrev4-1}
\bibliography{refs}

\end{document}